\documentclass[prl,twocolumn,showpacs,amsmath,amssymb,floatfix]{revtex4}
\usepackage{graphicx}
\usepackage{dcolumn}
\usepackage{bm}

\begin{document}

\title{Coherent-feedback quantum control with a dynamic compensator}
\author{Hideo Mabuchi} \email{hmabuchi@stanford.edu}
\affiliation{Physical Measurement and Control, Edward L.\ Ginzton Laboratory, Stanford University}

\date{March 12, 2008}
\pacs{02.30.Yy,42.50.-p,07.07.Tw}

\begin{abstract}
I present an experimental realization of a coherent-feedback control system that was recently proposed for testing basic principles of linear quantum stochastic control theory [M.~R.~James, H.~I.~Nurdin and I.~R.~Petersen, to appear in IEEE Transactions on Automatic Control (2008), arXiv:quant-ph/0703150v2]. For a dynamical plant consisting of an optical ring-resonator, I demonstrate $\sim 7$ dB broadband disturbance rejection of injected laser signals via all-optical feedback with a tailored dynamic compensator. Comparison of the results with a transfer function model pinpoints critical parameters that determine the coherent-feedback control system's performance.
\end{abstract}

\maketitle

\noindent The need for versatile methodology to control quantum dynamics arises in many areas of science and technology \cite{Mabu05}. For example, quantum dynamical phenomena are central to quantum information processing, magnetic resonance imaging and protein structure determination, atomic clocks, SQUID sensors, and many important chemical reactions. Substantial progress has been made over the past two decades in the development of intuitive approaches within specific application areas \cite{Peir88,Bela83,Huan83,Wise93a,Ahn02,Khan02,Jame04,Koro99} but the formulation of an integrated, first-principles discipline of quantum control---as a proper extension of classical control theory---remains a broad priority.

In our contemporary view it is natural to distinguish among three basic modes of quantum control: {\it open-loop}, in which a quantum system is driven via some time-dependent control Hamiltonian in a pre-determined way; {\it measurement-feedback}, in which discrete or continuous measurements of some output channel of an open quantum system are used to adjust the control actions in real time; and {\it coherent-feedback}, in which a quantized field scattered by the quantum system of interest is processed coherently (without measurement) and then redirected into the system in order to effect control. The first two modes are entirely analogous with classical open-loop and real-time feedback control, and their relation to existing engineering theory is now well understood \cite{Mabu05}. The possibility of coherent feedback, however, gives rise to a genuinely new category of control-theoretic problems as it encompasses non-commutative signals and quantum-dynamical transformations thereof \cite{fn1}. While some intriguing proposals can be found in the physics literature \cite{Wise94,Sher06}, relatively little is yet known about the systematic control theory of coherent feedback \cite{fn2}.

This article describes an experimental implementation of coherent-feedback quantum control with optical resonators as the dynamical systems and laser beams as the coherent disturbance and feedback signals. It is presented in the context of recent developments in control theory \cite{Jame07a,Jame07b,Nurd07}, which have shown that optimal and robust design of quantum coherent-feedback loops can be accomplished (in certain settings) using sophisticated methods of systems engineering (the setup parallels the quantum-optical system analyzed in \cite{Jame07a}). From the perspective of quantum information science, the results presented here represent a first step towards the goal of developing embedded, autonomous controllers that can implement feedback protocols for error correction without ever bringing signals up to a classical, macroscopic level.

\begin{figure}[tb!]
\includegraphics[width=0.42\textwidth]{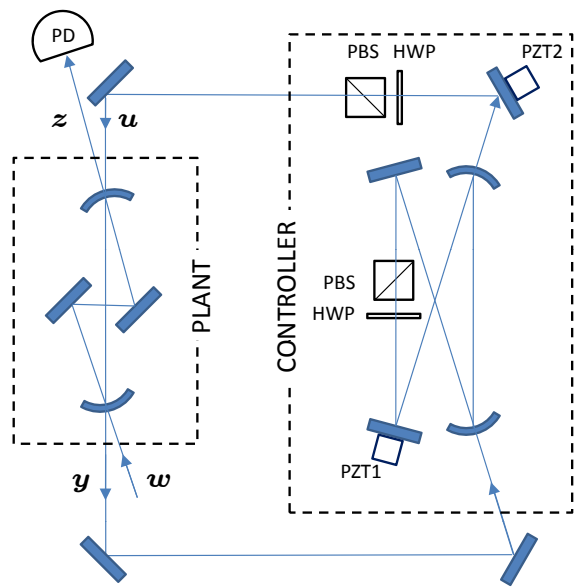}

\caption{\label{fig:sche} Schematic diagram of the experimental apparatus showing the coupled plant and controller resonators, variable optical attenuators (PBS/HWP), piezoelectric transducers (PZT) and photodetector (PD).}
\vspace{-0.1in}
\end{figure}

Fig.~\ref{fig:sche} presents a schematic overview of the apparatus and the coherent feedback loop. Two optical ring-resonators represent the ``plant'' and ``controller'' dynamical systems; the control-theoretic design goal is to tailor the properties of the controller so as to minimize the optical power detected at output $z$ when a ``noise'' signal (optical coherent state with arbitrary time-dependent complex amplitude) is injected at the input $w$. The component $y$ of the noise beam that reflects from the plant input coupler is treated as the error signal, which is coherently processed by the controller to produce a feedback signal $u$. The latter signal is fed back into the plant resonator via the output coupler, matched spatially to the same resonant mode driven by the noise input $w$.

While this type of coherent feedback loop is properly described using quantum stochastic differential equations (as in \cite{Jame07a}), a simplified analysis can be performed using Laplace transfer functions as familiar from classical linear control theory \cite{DFT}. This simplification here corresponds to the standard practice of tracking only the mean values of quantum observables, which is appropriate as long as all Hamiltonians are linear and all input states are gaussian. The open-loop (without feedback) transfer function of the plant resonator from the input $w$ to the output $z$ can be written $G_{zw}=-2\sqrt{k_1k_4}/(s+\gamma_p)$, where $\gamma_p$ is the total plant decay rate, $k_1$ and $k_4$ are the partial rates associated with transmission through the input and output couplers, and $s$ is the Laplace transform variable shifted to have value zero at the plant resonance frequency. In terms of physical parameters $\gamma_p=c\left( t_1^2 + t_2^2 + t_3^2 + t_4^2 + l^2\right)/4\pi L_p$, where $c$ is the speed of light, $L_p$ is the round-trip cavity length, $t_i^2$ is the power transmission coefficient of the $i$th mirror and $l^2$ represents all other intracavity losses. When the feedback loop is implemented as shown in Fig.~\ref{fig:sche} with a controller having transfer function $K_{uy}$, the overall (closed-loop) transfer function from $w$ to $z$ becomes
\begin{equation}
\text{S}\left( G,K\right)=G_{zw} + G_{zu}(1-K_{uy}G_{yu})^{-1}K_{uy}G_{yw},
\end{equation}
where the additional transfer functions are given for our setup by $G_{yu}=G_{zw}$, $G_{zu}=1-2k_4/(s+\gamma_p)$ and $G_{yw}=1-2k_1/(s+\gamma_p)$.

The {\it disturbance rejection} problem can now be defined as that of designing the controller so as to minimize the magnitude of $\text{S}\left( G,K\right)$. This corresponds to tailoring the coherent-feedback loop in such a way as to minimize the ratio of the optical power in the output $z$ to that of the noise input $w$. If we are interested in broadband disturbance rejection, it is important to note that $\text{S}\left( G,K\right)$ cannot be made much smaller than $G_{zw}$ for all values of $s$ if $K_{uy}$ is independent of $s$. Thus a simple proportional controller (for which $K_{uy}$ is a complex number) will not suffice; we require a dynamic compensator with non-trivial frequency response.

If we assume that our controller is itself an optical resonator, and with some foresight parametrize $K_{uy}$ as
\begin{equation}
K_{uy} = \frac{2\sqrt{\eta_K}\sqrt{k_1k_4}}{s+\gamma_p-2(k_1+k_4)+\eta_\gamma},
\end{equation}
it follows that $\text{S}\left( G,K\right)\rightarrow 0$ for all $s$ as $\eta_\gamma\rightarrow 0$ and $\eta_K\rightarrow 1$. Under these ideal conditions, zero optical power would be observed by a perfect photodetector monitoring the output $z$, for any coherent optical noise signal (mode-matched laser beam with arbitrary time-varying complex amplitude) injected at the input $w$. In practice it is difficult to implement the ideal controller transfer function exactly, and there is an additional challenge of perfecting the spatial mode-matching from the controller output to the $u$ input of the plant cavity. In modeling the experiment I thus include a factor $\mu$ to account for imperfect spatial mode-matching, $\eta_K$ to represent deviation of the magnitude of $K_{uy}$ from its ideal value, and $\eta_\gamma$ to represent deviation of the controller decay rate from its ideal value of $\gamma_p-2k_1-2k_4$. Writing $\text{S}_\mu\left( G,K\right)$ to denote the inclusion of a mode-matching correction, the ratio of the optical power in $z$ in the closed-loop case to the open-loop case is then
\begin{equation}
\left\vert\frac{\text{S}_\mu\left( G,K\right)}{G_{zw}}\right\vert ^2=\vert 1\pm
\sqrt{\mu}S_m\vert^2+(1-\mu)\vert S_u\vert^2,
\label{eq:Pz}
\end{equation}
(with $+$ for negative and $-$ for positive coherent feedback) where
\begin{eqnarray*}
S_m&=&G_{zw}^{-1}G_{zu}\left(1-\sqrt{\mu}K_{uy}G_{yu}\right)^{-1}K_{uy}G_{yw},\\
S_u&=&G_{zw}^{-1}K_{uy}G_{yw}.
\end{eqnarray*}
With reference to Fig.~\ref{fig:sche}, note that the resonance frequency of the controller cavity is adjustable via the actuator PZT1 and that the phase of $K_{uy}$ is continuously variable via PZT2. In practice these must both be set appropriately in order to minimize the magnitude of $\text{S}_\mu\left( G,K\right)$.

In my experiment the plant cavity is a four-mirror folded ring resonator (as depicted in Fig.~\ref{fig:sche}) with measured values $\gamma_p\approx 9.3$ MHz and $L_p\approx 14.1$ cm. The controller is a four-mirror ring resonator with measured decay rate $\gamma_c\approx 7.3$ MHz and length $L_c\approx 48.6$ cm. The controller decay rate can be adjusted using the intracavity variable attenuator shown in Fig.~\ref{fig:sche}. The injected signal at $w$, and thus the coherent-feedback loop signals $y$, $u$ and $z$, derive from an 852nm diode laser. The photodetector monitoring the output signal $z$ is placed behind an 852nm optical-bandpass filter. Additional laser beams from an 894nm diode laser are injected into both cavities in order to match the controller resonance frequency with that of the plant. The carrier frequency of the 894nm laser is locked to the plant cavity resonance (which is allowed to drift freely); PZT1 is then used to lock the controller cavity resonance to an electrooptic sideband of the 894nm laser that can be tuned over a frequency range greater than the controller cavity free spectral range. Using this arrangement it is straightforward to servo-control the controller cavity length so that its resonance frequency coincides with that of the plant cavity for the 852nm signal beams.

\begin{figure}[tb!]
\includegraphics[width=0.47\textwidth]{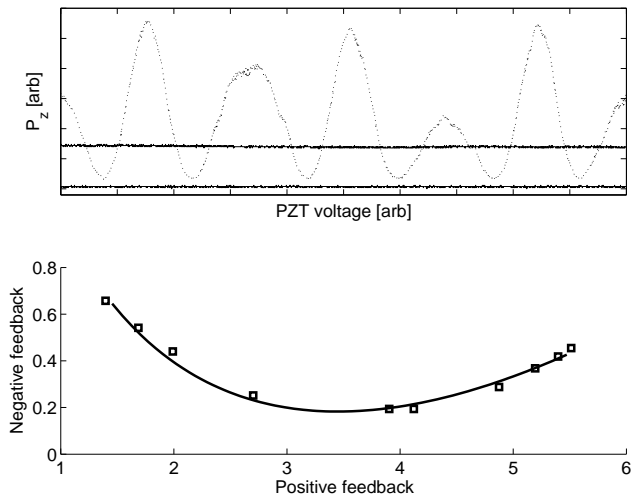}
\caption{\label{fig:phase} Upper: variation of the total optical power in output $z$ as PZT2 is scanned to change the coherent-feedback phase. Lower: parametric plot of the $s=0$ closed-loop system response under positive (x-coordinate) and negative (y-coordinate) feedback with gain mismatch $\eta_K$ ranging between approximately 0.06 and 2.2; see text for an explanation of the theoretical curve.}
\vspace{-0.1in}
\end{figure}

Disturbance rejection via coherent feedback is demonstrated in Fig.~\ref{fig:phase}. In order to generate this data the 852nm laser is servo-controlled to track the resonance frequency of the plant cavity, thus ensuring that we initially probe all transfer functions with $s=0$. In the upper subplot, the two solid traces show the electronic noise floor (lower) and the optical power level detected at the photodetector (PD) monitoring the output $z$ in the absence of coherent feedback (upper). When the coherent-feedback loop is closed, the optical power in $z$ is seen to depend strongly on the coherent feedback phase as set by PZT2. If the voltage on PZT2 is ramped in order to vary the phase continuously the coherent-feedback loop oscillates between positive and negative feedback (dashed trace). The ratio of the minimum value of the optical power in $z$, obtained with negative coherent feedback, to the open-loop value yields an optimal disturbance rejection of approximately 7 dB. As the overall system is linear, the input power of $w$ is unimportant and drops out of the analysis; values in the range of $\sim 100$ $\mu$W were used in this experiment.

The lower subplot of Fig.~\ref{fig:phase} presents a parametric plot of the maximum (horizontal axis) versus minimum (vertical axis) optical power ratios observed with $\eta_K$ ranging over a set of values between 0.06 and 2.2 (adjusted using the variable attenuator at the output of the controller cavity). The curve shows the prediction obtained from the transfer-function model described above, where the values of $\eta_\gamma$, $\mu$, $k_1$, and $k_4$ were adjusted within reasonable ranges to obtain a good fit to the data. The values obtained in this way are $\eta_\gamma=\eta_p/14$, $\mu=0.84$ and $t_1^2=t_4^2=0.002$. These values for the plant input- and output-coupler power transmission coefficients agree with witness sample measurements from the mirror coating run when adjusted for the beam incidence angle of $0.3$ radians. The mode-matching factor $\mu$ is in agreement with a direct measurement $\mu\alt 0.85$ obtained by observing the ratio of TEM00 to transverse-mode transmission peaks, and $\eta_\gamma$ agrees with the measurements of $\gamma_p$ and $\gamma_c$ when $t_1^2$ and $t_4^2$ are set to 0.002. This comparison of data with the transfer-function model thus confirms that the parameters affecting system performance are well known; it follows from Eq.~(\ref{eq:Pz}) that the $s=0$ disturbance rejection performance is fundamentally limited by imperfect mode-matching $\mu<1$. The model also shows that for low-frequency noise (small detuning), a small error $\eta_\gamma$ in the controller decay rate can be overcome by adjusting $\eta_K$. But a high degree of broadband disturbance rejection (for noise inputs with frequency spread comparable to $\gamma_p$) is not possible unless $\eta_\gamma$ is carefully minimized.

\begin{figure}[tb!]
\begin{centering}
\includegraphics[width=0.47\textwidth]{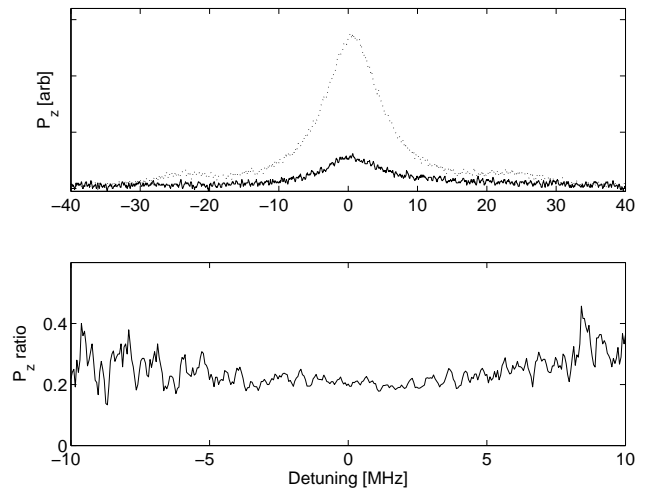}
\end{centering}
\caption{\label{fig:ss} Upper: optical power in $z$ as a function of detuning between the input laser frequency and the plant resonance, corresponding to swept sine measurements of $|G_{zw}|^2$ (dashed) and $|\text{S}_\mu(G,K)|^2$ (solid). Shoulders on the resonance peak are electrooptic sidebands that were added to establish the frequency scale. Lower: ratio of traces from the upper plot, confirming the broadband nature of the disturbance rejection.}
\vspace{-0.1in}
\end{figure}

The upper subplot of Fig.~\ref{fig:ss} displays data corresponding to (single shot) swept-sine transfer function measurements of $G_{zw}$ (dashed) and $\text{S}_\mu\left( G,K\right)$ under negative coherent feedback (solid). These were obtained simply by scanning the 852nm laser frequency over the plant cavity resonance and recording the optical power in $z$. The ratio of the two traces is shown in the lower subplot, which shows that suppression is achieved over a wide range of noise signal frequencies and thus establishes the broadband nature of the disturbance rejection.

\begin{figure}[tb!]
\includegraphics[height=2.5in]{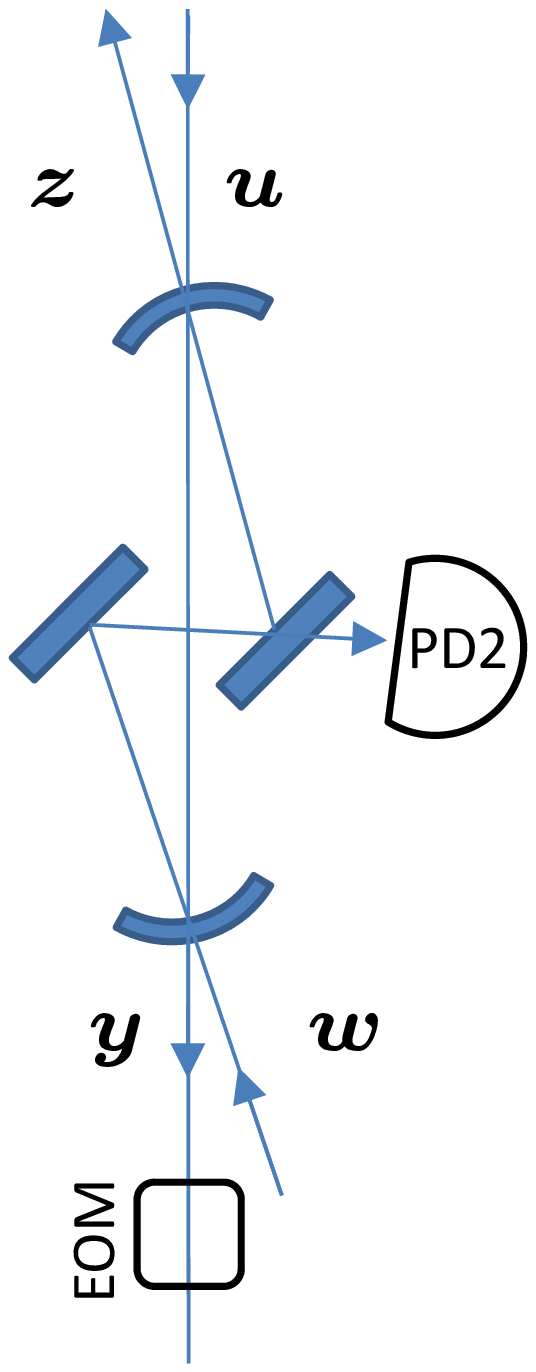}
\hspace{0.25in}
\includegraphics[width=0.2\textwidth]{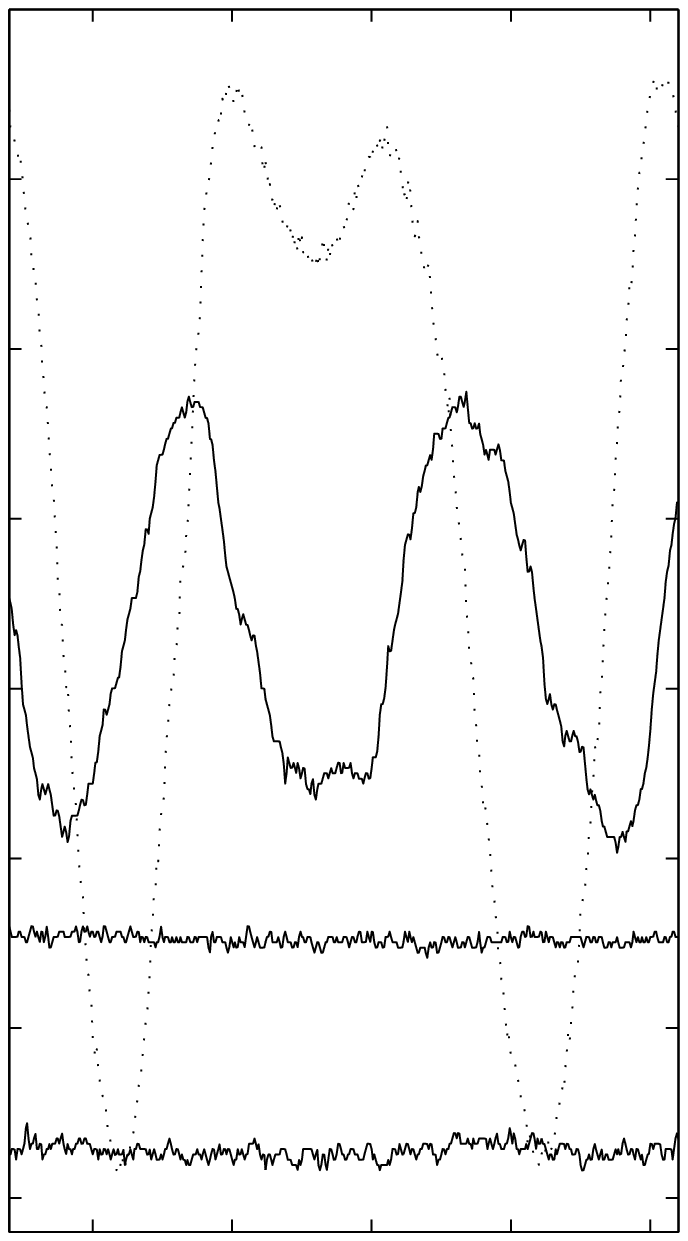}
\caption{\label{fig:lock} Left: schematic detail of the plant indicating additional photodetector (PD2) and electrooptic modulator (EOM) used to generate an error signal for locking the coherent-feedback phase. Right: oscillating traces depict the optical power variation of output $z$ (dashed) and the coherent-feedback phase error signal (solid) as a function of coherent-feedback phase. The flat traces indicate the levels of optical power observed in output $z$ with no feedback (upper) and with the phase locked to the negative feedback condition (lower).}
\vspace{-0.1in}
\end{figure}

Finally, Fig.~\ref{fig:lock} shows that it is possible to stabilize the path-length of the coherent signal loop to maintain negative feedback. In order to obtain a suitable error signal an electrooptic modulator (EOM) is inserted in the $y$ signal path as shown in the left-hand diagram. This EOM is driven by a high-frequency ($\sim 1$ MHz) sine wave; the signal from a photodetector (PD2) at an auxiliary output port of the plant cavity is demodulated at this frequency to produce the error signal. On the right side of Fig.~\ref{fig:lock}, the oscillating traces depict the variation of the closed-loop output power in $z$ (dashed) and the error signal (solid) as the length of the feedback loop is varied using PZT2. The minimum of the closed-loop output power is seen to coincide with a zero-crossing of the error signal, thus making it possible to stabilize the coherent-feedback phase via electronic feedback to PZT2. The open-loop and phase-stabilized closed-loop optical output powers in $z$ (with the 852nm laser locked to the plant cavity resonance) are indicated by the flat traces in the plot.

While the experiment presented here has dealt only with coherent optical states, the coherent-feedback disturbance rejection scheme should function without significant modification for a very broad class of quantum noise signals. Existing theory based on quantum stochastic differential equations provides a rigorous basis for predicting the performance expected for squeezed-state inputs, and indeed the type of dynamic compensation demonstrated here is already of interest for tailoring spectral properties of squeezed light for applications such as gravity-wave detection \cite{SS}. From an experimental perspective it would be most interesting to test the performance with non-gaussian quantum states such as those produced by photon-subtraction \cite{NGS}, which would push beyond the reach of current theory.

\begin{acknowledgments}
This work was supported by the Air Force Office of Scientific Research under grant number FA9550-07-1-0198.
\end{acknowledgments}


\begin{thebibliography}{99}
\bibitem{fn1} Optical feedback is already a common technique in applications such as diode laser stabilization \cite{Lang80} and photonic signal processing \cite{Clav07}; it is likewise widely exploited in basic studies \cite{Gutl05} and applications \cite{Liu01} of nonlinear dynamics and chaos. The present work is distinguished by its ties to recent developments in the theory of open quantum systems and quantum control.
\end{thebibliography}
\end{document}